\documentclass[12pkt]{article}
\usepackage[margin=2cm]{geometry}
\usepackage{amsmath}
\usepackage{amssymb}
\usepackage{xfrac}
\usepackage{soul}

\newcommand{\calO}{\mathcal{O}}

\newcommand{\als}{\alpha_S}

\newcommand{\ptt}{\ensuremath{p_T}}
\newcommand{\mh}{m_{H}}

\newcommand{\muR}{\mu_{R}}
\newcommand{\muF}{\mu_{F}}

\title{Effective Field Theory for Higgs properties parametrisation: the transverse momentum spectrum case}

\author{Massimiliano Grazzini$^a$, \underline{Agnieszka Ilnicka}$^{abc}$, Michael Spira$^c$, Marius Wiesemann$^{ad}$\\
\small $^a$ Physics Institute, University of Z\"urich, Winterthurerstrasse 190, CH-8057 Z\"urich\\
\small $^b$ Physics Department, ETH Z\"urich, Otto-Stern-Weg 5, CH-8093 Z\"urich\\
\small $^c$ Paul Scherrer Institute, CH-5232 Villigen PSI\\
\small $^d$ CERN Theory Division, CH-1211, Geneva 23\\
\small grazzini@physik.uzh.ch, ailnicka@physik.uzh.ch, michael.spira@psi.ch, marius.wiesemann@cern.ch}
\date{\vspace{-5ex}}
\begin{document}
\maketitle
\begin{abstract}
After the Higgs boson discovery, LHC can be used as a precision machine to explore its properties. Indeed, in case new resonances will not be found, the only access to New Physics would be via measuring small deviations from the SM predictions. A consistent approach is provided by a bottom-up Effective Field Theory, with dimension six operators built of Standard Model fields (SMEFT). We discuss how this approach works in case of the transverse momentum spectrum of the Higgs particle. In our calculation we augmented the Standard Model with three additional operators describing modifications of the top and bottom Yukawa couplings, and a point-like Higgs coupling to gluons. Based on recently presented resummed transverse-momentum spectra including these operators at NLL+NLO accuracy,which 
we recently presented in Ref \cite{ours}, in this note we show the extension of the calculation to NNLL+NNLO. We find that such modifications, while affecting the total rate within the current uncertainties, can lead to significant distortions of the spectrum. Additionally, through significant reduction of the scale uncertainty we improve the sensitivity to the new operators.
\end{abstract}

\section{Introduction}

The 125 GeV scalar resonance observed in ATLAS and CMS at the LHC in 2012 \cite{ATLASdisc,CMSdisc} closely resembles the Higgs boson of the Standard Model (SM).
It is however well known that the SM cannot predict phenomena such as the baryon asymmetry of the Universe or dark matter. 
Many beyond the SM (BSM) theories addressing these issues have been developed, which in particular often modify the Higgs boson properties. It is possible that new resonances exist beyond the reach of LHC, and New Physics will be manifested just by small deviations from the SM predictions. Effective Field Theory (EFT), a consistent way to parametrise these deviations, is an approach in which the unknown high-scale fields are integrated out and leave an infinite ladder of 
dimension higher than 4 operators, with a well-defined hierarchy. The EFT can thus be used to build a bottom-up 
approach in which the usual dimension-four operators of the SM are augmented by 
leading (dimension-six) operators\footnote{The full set of dimension-six SMEFT operators has been presented in
\cite{dim61,dim62}.}. With the use of experimental data, the values of the Wilson coefficients of these operators can then be fixed. For this, however, tools including dimension-six operators need to be developed. Since the Higgs \ptt{} spectrum is an important observable, which will be measured at the LHC, that can shed light on Higgs' properties, the aim is to develop a dedicated tool.

The inclusion of dimension-six and dimension-eight operators in the $p_T$-spectrum has been considered in Refs.~\cite{ptdim61,ptdim62,ptdim63,Maltoni:2016yxb} and \cite{ptdim81,ptdim82}, respectively.
Strategies for extracting information on the Higgs-gluon coupling from the measurements were studied in Ref.~\cite{ptdim63}. Most of the above studies, however, are limited to the high-$p_T$ region of the spectrum, and do not include
small-$p_T$ resummation. In particular, in our study, we employ analytic resummation to deal with the low-$p_T$ region. In this contribution we present the resummed $p_T$-spectrum at NNLL+NNLO accuracy, with the inclusion of a set of dimension-six parameters relevant for Higgs boson production.

\section{Effective operators}

We consider the effective Lagrangian
\begin{equation}
{\cal L}={\cal L}_{SM}+\sum_i \frac{c_i}{\Lambda^2}{\cal O}_i
\end{equation}
where the SM is supplemented by the inclusion of a set of dimension-six
operators describing new physics effects at a scale $\Lambda$ well above
the electroweak scale.  In the spectra presented below we included three additional operators:
\begin{equation}
{\cal O}_1 = |H|^2 G^a_{\mu\nu}G^{a,\mu\nu}\,,\quad {\cal O}_2 = |H|^2 \bar{Q}_L H^c u_R + h.c.\,, \quad {\cal O}_3 = |H|^2 \bar{Q}_L H d_R + h.c.
\end{equation}
These operators, in the case of single Higgs production, may be expanded as:
\begin{equation}
\frac{c_1}{\Lambda^2}\,{\cal O}_1 \rightarrow \frac{\als}{\pi v} c_g h  G^a_{\mu\nu}G^{a,\mu\nu}\,, \quad
\frac{c_2}{\Lambda^2}\,{\cal O}_2 \rightarrow \frac{m_t}{v} c_{t} h \bar{t} t\,, \quad
\frac{c_3}{\Lambda^2}\,{\cal O}_3 \rightarrow \frac{m_b}{v} c_{b} h \bar{b} b\,,
\end{equation}

The operator ${\cal O}_1$ corresponds to a contact interaction between
the Higgs boson and gluons with the same tensor structure as in the heavy-top
limit (HTL) of the SM. The operators ${\cal O}_2$ and  ${\cal O}_3$ describe
modifications of the top and bottom Yukawa couplings. In our
convention, based on the SILH basis \cite{SILH0,SILH}, we express the
Wilson coefficients as factors in the canonically normalized Lagrangian. 

The coefficients $c_t, c_b$ and $c_g$ can be probed in Higgs boson processes.
In particular, $c_t$ (and $c_b$) may be measured in the $t\bar{t}H$ (and
$b\bar{b}H$) production modes. The
coefficient $c_b$ can also be accessed through the decay $H \rightarrow
b\bar{b}$.
\section{Calculation Setup}

The basis for the calculation presented here are the NLL+NLO spectra of ref. \cite{ours}. The calculation relies on the codes of Refs. \cite{hqt1,hqt2} and \cite{Harlander:2014uea,aprmt1}. To provide the state-of-the-art NNLL+NNLO SMEFT calculations, we used the NNLL+NNLO SM predictions, and then scaled them with the NLL+NLO calculations including the SMEFT operators:
\begin{equation}
\left(\frac{d \sigma}{d p_t}\right)_{NNLL+NNLO}^{SMEFT}(p_T) = \frac{\left(\frac{d \sigma}{d p_t}\right)_{NLL+NLO}^{SMEFT} (p_T)}{\left(\frac{d \sigma}{d p_t}\right)_{NLL+NLO}^{SM} (p_T)}\cdot \left(\frac{d \sigma}{d p_t}\right)_{NNLL+NNLO}^{SM}(p_T)
\end{equation}
It is important to note here, that the NNLL+NNLO results \cite{ptNLO1,ptNLO2,ptNLO3} are known only in the heavy top limit, with just approximate inclusion of finite top mass effects. We used SM results obtained with the numerical code HRes \cite{HRes1,HRes2}, and we ensured the set up as close as possible to the one used in the NLL+NLO calculations. \footnote{See \cite{ours} for details.}

To estimate the uncertainties due to the renormalization and factorization scales we performed the customary seven-point
$\muR$, $\muF$ variation, i.\,e. we consider independent variations
within the range $\mu_0/2 \le\muF,\muR\le 2\mu_0$ with
$1/2<\muR/\muF<2$, where $\mu_0=\sqrt{p_T^2+\mh^2}/2$. We then varied also the two resummation scales
by a factor of two. \footnote{Note that here only two resummation scales were used, corresponding to the top and interference ones ($Q_t$ and $Q_{int}$) from previous study. }

\section{Results}

\begin{figure}[h!]
\begin{center}
\begin{tabular}{cc}
\hspace*{-0.17cm}
\includegraphics[trim = 7mm -7mm 0mm 0mm, width=.36\textheight]{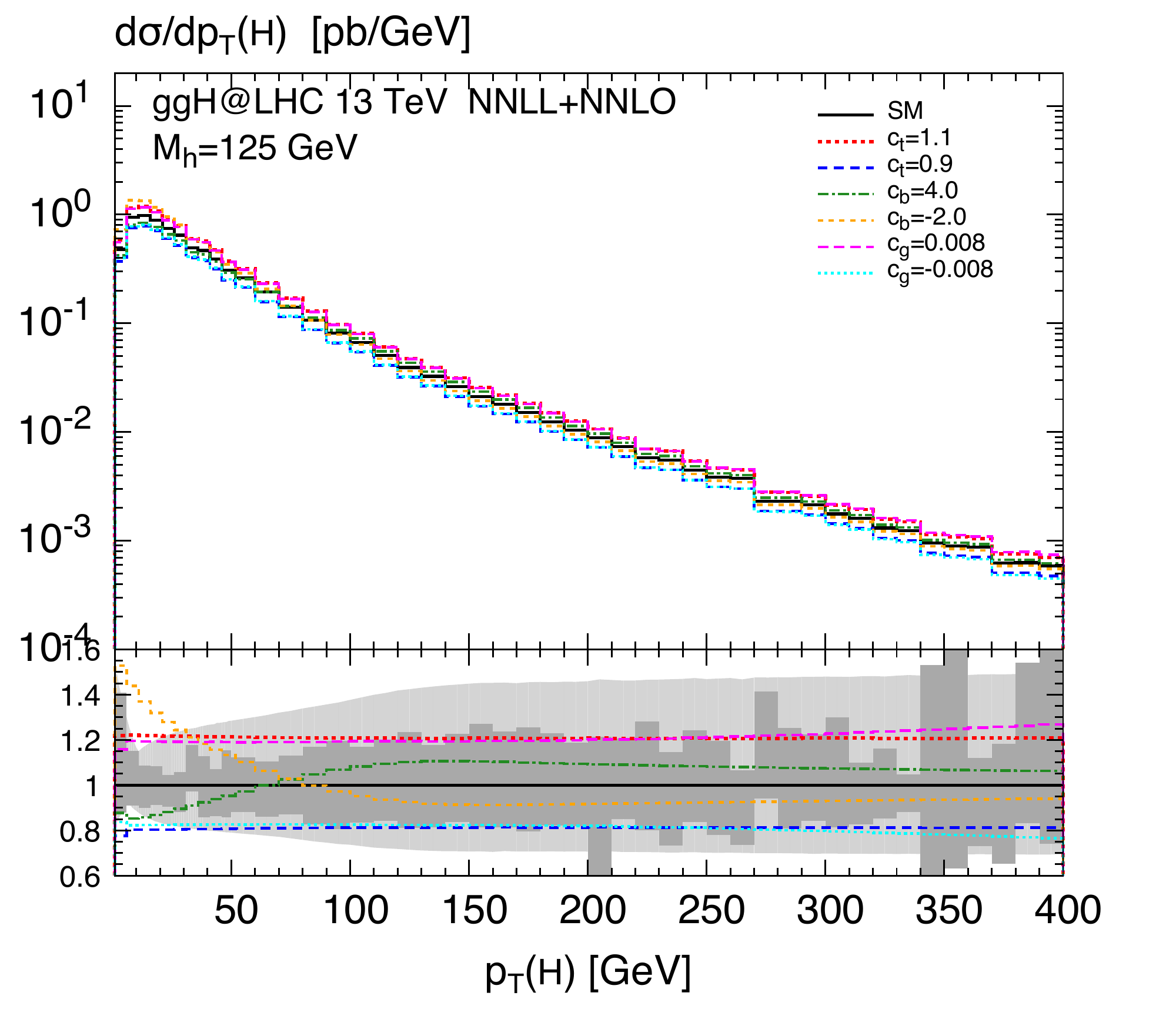} &
\includegraphics[trim = 7mm -7mm 0mm 0mm, width=.36\textheight]{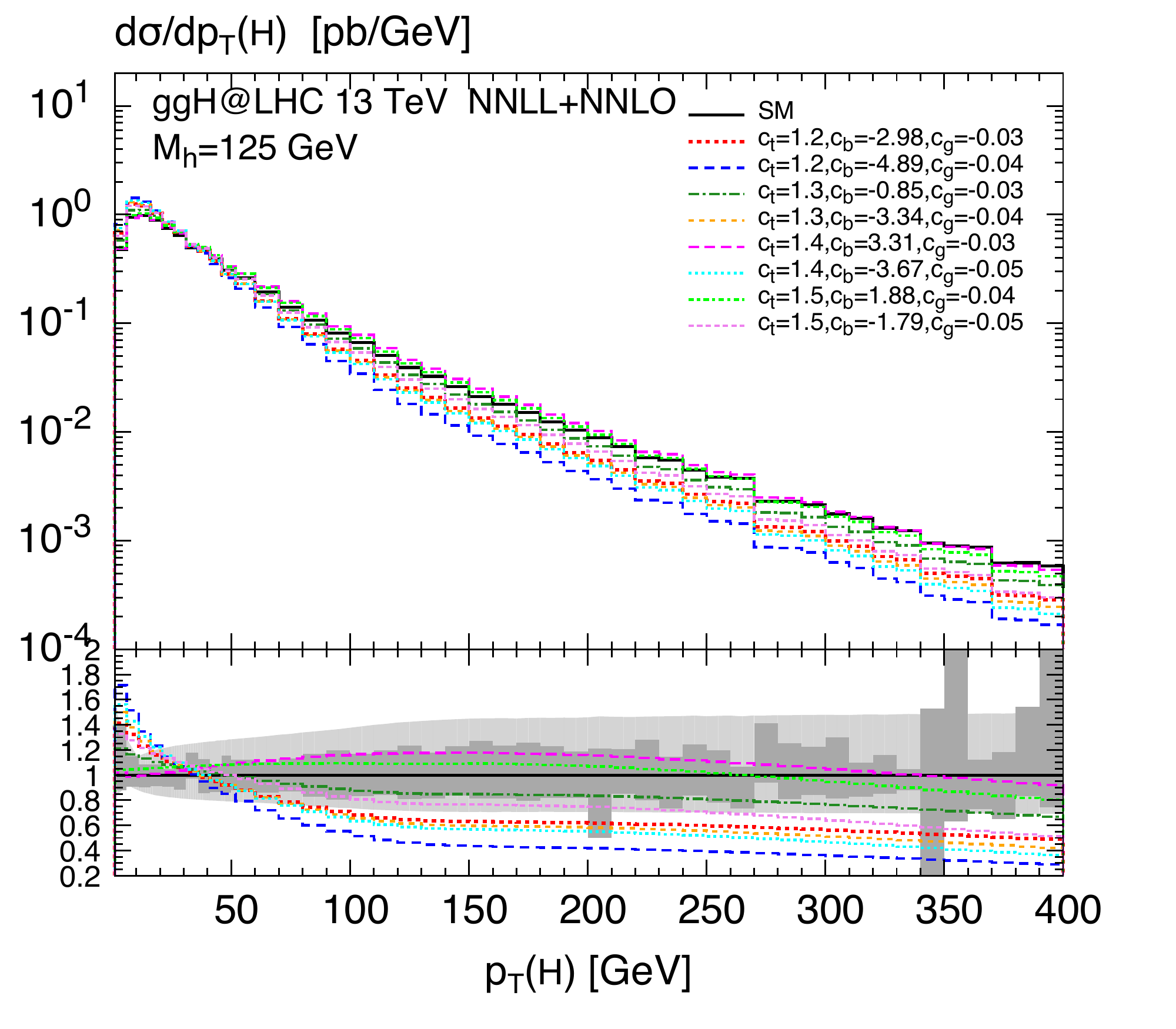} \\[-1em]
(a) & (b)
\end{tabular}
\caption[]{\label{fig}{Higgs transverse-momentum spectrum in the SM 
(black, solid) compared to  (a) separate variations and (b) mixed contribution of the dimension-six operator
for 0 GeV$\le p_T \le400$ GeV. 
The lower frame shows the ratio with respect to the SM prediction.
The shaded lighter and darker grey bands in the ratio indicates the uncertainty due to scale variations in NLL+NLO and NNLL+NNLO case respectively. 
See text for more details.}}
\end{center}
\end{figure}

In this section we present the $p_T$ spectra of the Higgs particle, including the modifications coming from the effective operators. The values used for the Wilson coefficients correspond to Figures 3 and 6 in \cite{ours}. \footnote{The procedure described above can be redone also for all the other operator combinations.} We note, that all presented spectra correspond to a total cross section within 20\% deviations from the SM prediction. As reference the NNLL+NNLO SM predictions are shown in the Figures as solid black line, with the grey bands on the lower panel showing the perturbative uncertainties. The light grey band correspond to the NLL+NLO uncertainty relative to the central scale NLL+NLO calculations, while the darker band corresponds to the NNLL+NNLO scale uncertainty relative to the NNLL+NNLO central scale calculation. This presentation allows us to observe the decrease of the uncertainty while going one order higher. It results in the decrease of uncertainty by about a factor of two in the low and intermediate $p_T$ range (up to around 250 GeV). In the higher \ptt{} region, the uncertainties become more scattered due to statistical fluctuations.  

In Figure \ref{fig} (a) we present the spectrum with the individual contributions of the operators. From the grey bands on the lower panel it can be noticed that at the NLL+NLO accuracy all the curves are within the scale uncertainty, while in the NNLL+NNLO case the effects of higher dimension operators exceed the uncertainties. As already noticed in \cite{ours}, the encouraging fact is that the modifications of different operators manifest themselves mostly in different regions of \ptt{}, i.e. $c_b$ in low and $c_g$ in high \ptt{} regions.

The spectra presented in Figure \ref{fig} (b) correspond to switching on all three SMEFT operators. We choose scenarios with increased top-quark Yukawa coupling (up to $c_t=1.5$), as hinted by the excess on the $t\bar{t}H$ rate over the SM prediction reported in ATLAS and CMS \cite{ATLAS:2016awy,CMS:2016vqb}. As it was noticed also in the NLL+NLO case most of the scenarios distort the shape of the spectra beyond the scale uncertainty, but the further reduction of the scale uncertainty in the NNLL+NNLO case allows also for a better discrimination between different scenarios. \footnote{For more discussion on the SMEFT operators impact on the spectra refer to \cite{ours}.}

\section{Conclusions}

If New Physics will not be accessible at the LHC through direct searches, e.g., with
the discovery of new resonances, it will be crucial to fully exploit the data
to study possible (small) deviations from the SM predictions. The formalism that can be used for this purpose is SMEFT, which parametrises high-scale BSM effects through appropriate higher-dimensional operators.
Bounds on the corresponding Wilson coefficients of these operators can be set by comparing to the experimental data.

In this note we have presented an extension of the recently published NLL+NLO calculations of the Higgs \ptt{} spectra augmented with SMEFT operators \cite{ours} to NNLL+NNLO level of accuracy. We start with state-of-the-art SM predictions and scale them by relative SMEFT/SM effects at NLL+NLO (i.e. the ratios plotted in the lower panels of the Figures). 

We found that variations of different SMEFT operators  
manifest themselves in different regions of the Higgs \ptt{} spectrum:  a modification of the bottom 
Yukawa coupling (${\calO}_3$) induces effects almost exclusively at small \ptt{}, while a direct coupling of the Higgs boson to gluons (${\calO}_1$) changes the shape of the 
distribution  in the high-$p_T$ tail and the top Yukawa coupling primary affects the normalisation. We notice from the presented spectra that the shape of the transverse momentum distribution depends on the mass of the particle that mediates the Higgs-gluon
coupling. The lower the mass of that particle, the softer is the resulting spectrum, and thus the enhancement of bottom loop leads to the softest spectrum, while an enhancement of the point-like coupling (corresponding to infinite mass particles in the loop) to the hardest one.

Finally we mention the limitation of our study. The NNLL+NNLO SM predictions are known only in the heavy top limit, with just approximate inclusion of top mass effects, and thus the approach involving a scaling of the spectra was the only possible. With the full top mass dependent results at NNLO it would be desirable to redo the analysis in the same spirit as the one done in \cite{ours} in the NLL+NLO case. Moreover, in our study we omitted the chromomagnetic operator, which can also have a relevant impact on the Higgs \ptt{} spectrum. This calculation is planned for future work. 

\section*{Acknowledgments}

AI thanks H. Sargsyan for support with the HRes program. This work is supported by the 7th Framework Programme of the European Commission through the Initial Training Network HiggsTools PITN-GA-2012-316704.

\bibliographystyle{iopart-num}
\bibliography{pteft}

\end{document}